\documentclass[aps,pra,12pt,amsmath,onecolumn,floatfix,showpacs]{revtex4}
\usepackage{amsmath}
\usepackage{amsfonts}
\usepackage{graphicx}

\newcommand{\real}{\operatorname{Re}}
\newcommand{\imag}{\operatorname{Im}}
\newcommand{\parti}[2]{\frac{\partial #1}{\partial #2}}
\newcommand{\partit}[2]{\frac{\partial^2 #1}{\partial #2^2}}

\newcommand{\ket}[1]{|#1\rangle}

\newcommand{\bra}[1]{\langle#1|}

\newcommand{\avg}[1]{\langle#1\rangle}
\newcommand{\Avg}[1]{\left\langle#1\right\rangle}

\newcommand{\abs}[1]{\left|#1\right|}
\newcommand{\bk}[1]{\left(#1\right)}
\newcommand{\Bk}[1]{\left[#1\right]}

\newcommand{\trace}{\operatorname{tr}}

\begin{document}

\title{Quantum Nonlocality in Weak-Thermal-Light Interferometry}

\author{Mankei Tsang}
\email{eletmk@nus.edu.sg}
\affiliation{Department of Electrical and Computer Engineering,
  National University of Singapore, 4 Engineering Drive 3, Singapore
  117583}

\affiliation{Department of Physics, National University of Singapore,
  2 Science Drive 3, Singapore 117551}

\affiliation{Center for Quantum Information and
  Control, University of New Mexico, MSC07--4220, Albuquerque, New
  Mexico 87131-0001, USA}


\date{\today}

\begin{abstract}
  In astronomy, interferometry of light collected by separate
  telescopes is often performed by physically bringing the optical
  paths together in the form of Young's double-slit
  experiment. Optical loss severely limits the efficiency of this
  so-called direct detection method, motivating the fundamental
  question of whether one can achieve a comparable performance using
  separate optical measurements at the two telescopes before combining
  the measurement results. Using quantum mechanics and estimation
  theory, here I show that any such spatially local measurement
  scheme, such as heterodyne detection, is fundamentally inferior to
  coherently nonlocal measurements, such as direct detection, for
  estimating the mutual coherence of bipartite thermal light when the
  average photon flux is low. This surprising result reveals an
  overlooked signature of quantum nonlocality in a classic optics
  experiment.
\end{abstract}

\maketitle

The basic goal of stellar interferometry is to retrieve astronomical
information from the mutual coherence between optical modes collected
by telescopes \cite{mandel,monnier,thompson}. The imaging resolution
increases with the distance between the collected optical modes called
the baseline, motivating the development of long-baseline stellar
interferometry using light collected from a telescope array
\cite{monnier,thompson}. The standard method of stellar interferometry
in the optical regime is called direct detection, which coherently
combines the optical paths in the form of the classic Young's
double-slit experiment, but its efficiency suffers from decoherence in
the form of accumulating optical loss along the paths as the baseline
is increased. To avoid optical loss, an alternative method is to
perform separate heterodyne detection at the two telescopes, before
combining the measurement results via classical communication and data
processing \cite{monnier,thompson}. In quantum information theory,
direct detection can be classified as a nonlocal measurement scheme,
which requires joint quantum operations on the two optical modes,
while heterodyne detection is a local measurement scheme, which does
not require quantum coherence between the separate detectors
\cite{holevo,peres}. Townes has previously analyzed the
quantum noises in direct and heterodyne detection and concluded that
direct detection is superior at high optical frequencies and
heterodyne detection is superior at low frequencies
\cite{thompson,townes}. Heterodyne detection is, however, only one
example of local measurements, and it remains a fundamental and
important question whether any other local measurement can perform as
well as nonlocal measurements while not suffering from decoherence.

The main purpose of this Letter is to prove that, in the case of weak
thermal light, \emph{any} local measurement scheme must be
significantly inferior to a nonlocal one for the estimation of the
mutual coherence according to quantum mechanics. This is a surprising
result in quantum metrology, since the disadvantage of local
measurements does not otherwise occur for coherent states at any
strength, a well-studied case in quantum metrology \cite{caves},
strong thermal light, in which case there is little difference between
direct and heterodyne detection \cite{sup}, or even the single-photon
state assumed by Gottesman, Jennewein, and Croke in their proposal of
shared-entanglement stellar interferometry \cite{gottesman}. This
quantum measurement nonlocality can be regarded as a dual of
Einstein-Podolsky-Rosen entanglement \cite{epr,holevo}: Despite the
fact that bipartite thermal light has a well-defined classical
description and possesses no quantum entanglement, nonlocal quantum
measurements are necessary to extract the most information from the
light. For optical interferometry and imaging applications in general,
the result demonstrates the fundamental advantage of nonlocal
measurements for weak thermal light and motivates the development of
coherent optical measurement techniques, such as integrated optical
information processing \cite{monnier,brady,psaltis} and entanglement
sharing \cite{gottesman}.

\begin{figure}[htbp]
\includegraphics[width=0.7\textwidth]{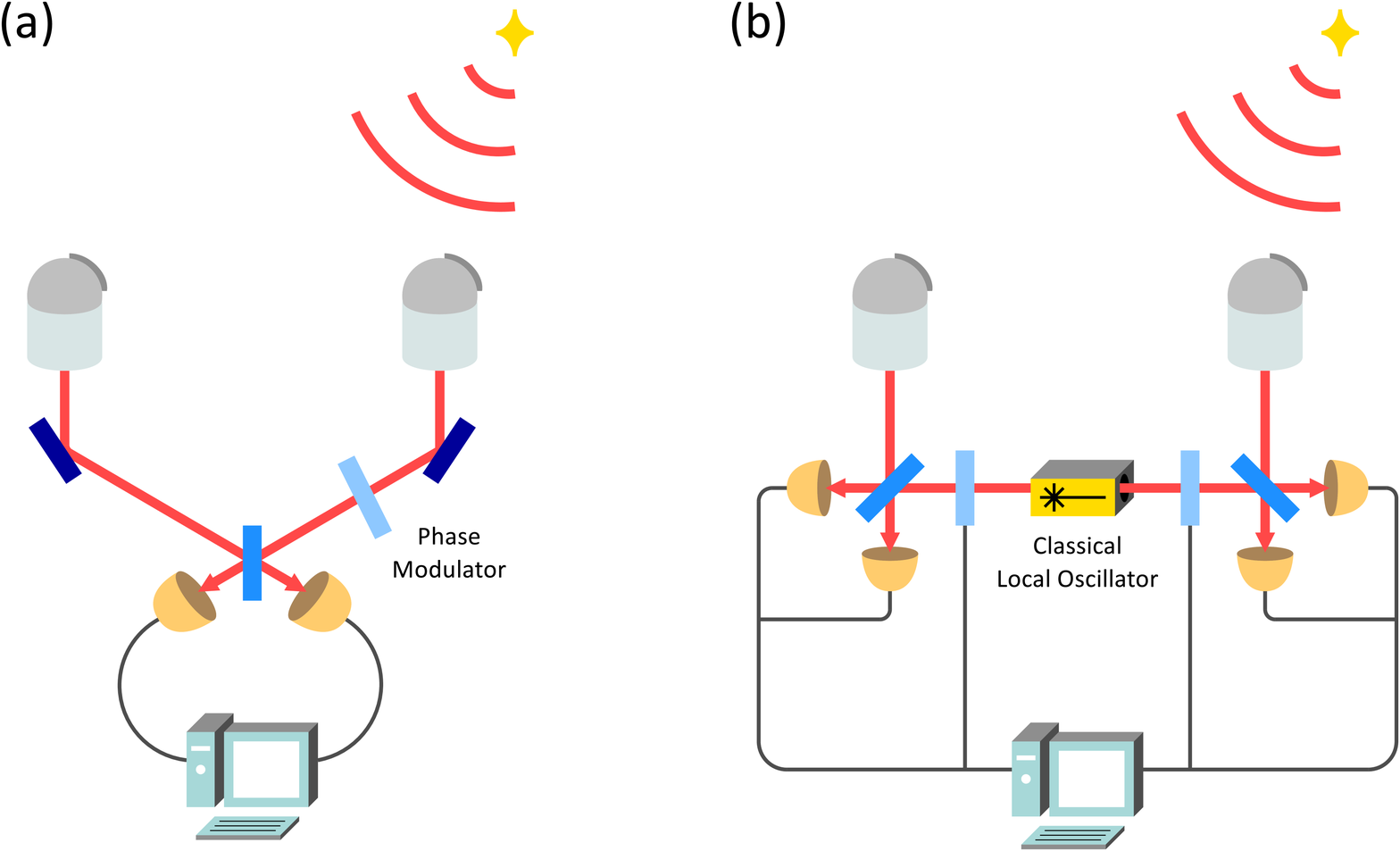}
\caption{(Color online). Schematics of (a) the direct detection
  scheme, an example of nonlocal quantum measurement, and (b) a local
  measurement scheme, which performs spatially separate measurements
  and permits only classical communication and control between the two
  sites. Examples of the latter include heterodyne and homodyne
  detection.}
\label{telescope}
\end{figure}

Consider the estimation of first-order spatial coherence ($g^{(1)}$)
between two distant optical modes. In quantum optics, bipartite
thermal light is described by the density operator
\begin{align}
\rho &= \int d^2\alpha d^2\beta 
\Phi(\alpha,\beta) \ket{\alpha,\beta}\bra{\alpha,\beta},
\label{thermal}
\end{align}
where $\ket{\alpha,\beta}$ is a coherent state with amplitudes
$\alpha$ and $\beta$ in the two modes and $\Phi(\alpha,\beta)$ is the
Sudarshan-Glauber representation \cite{mandel}, given by
\begin{align}
\Phi(\alpha,\beta) &= \frac{1}{\pi^2\det\Gamma}
\exp\Bk{-\bk{\begin{array}{cc}\alpha^* &\beta^*\end{array}}
\Gamma^{-1}
\bk{\begin{array}{c}\alpha \\\beta\end{array}}}.
\end{align}
$\Gamma$ is the mutual coherence matrix:
\begin{align}
\Gamma &\equiv 
\bk{\begin{array}{cc} \Gamma_{aa} & \Gamma_{ab}\\
\Gamma_{ba} & \Gamma_{bb}\end{array}}
= 
\bk{\begin{array}{cc}\avg{a^\dagger a} & \avg{b^\dagger a}\\
\avg{a^\dagger b} & \avg{b^\dagger b}\end{array}},
\end{align}
and $a$ and $b$ are annihilation operators of the optical modes.  The
zero-mean Gaussian statistics are a standard assumption for
astronomical sources in theoretical optics \cite{mandel,thompson}.
The positive $\Phi$ function indicates that the two modes are
classically correlated only and possess no quantum entanglement
\cite{werner}.

Let $\avg{a^\dagger a} = \avg{b^\dagger b} = \epsilon/2$ for
simplicity. For an incoming light with photon-flux spectral density
$S(\nu)$ and a relatively narrow detector bandwidth $\Delta\nu$ around
a center frequency $\nu_0$, the filtered photon flux is
$S(\nu_0)\Delta\nu$.  Over the duration of the effective temporal mode
$\Delta t\sim 1/\Delta\nu$, $\epsilon = S(\nu_0)\Delta\nu\Delta t\sim
S(\nu_0)$ turns out to be independent of the detector bandwidth and a
function of the source and the telescope efficiency only. Considering
the case $\epsilon\ll 1$, as is common for interferometry with high
optical $\nu_0$, the density operator can be approximated in the
photon-number basis as
\begin{align}
\rho &= (1-\epsilon)\ket{0,0}\bra{0,0}
+\frac{\epsilon}{2}
\big[\ket{0,1}\bra{0,1}+\ket{1,0}\bra{1,0}
\nonumber\\&\quad
+g^*\ket{0,1}\bra{1,0}+g\ket{1,0}\bra{0,1}\big]
+O(\epsilon^2),
\label{rho_number}
\end{align}
where I have defined $\epsilon g/2 \equiv \Gamma_{ab} = \Gamma_{ba}^*$
and $g \equiv g_1 + ig_2$ as the complex degree of coherence with $|g|
\le 1$ \cite{mandel}. In the following, I neglect the small
$O(\epsilon^2)$ terms, assume that $\epsilon$ is known, and $g_1$ and
$g_2$ are the unknown parameters to be estimated. The assumption of a
known $\epsilon$ should be reasonable, as other non-interferometric
imaging methods can be used to estimate the average photon flux and
are usually much less sensitive to noise \cite{monnier}. Otherwise
$\epsilon$ should also be regarded as an unknown parameter to be
estimated by the interferometer, a complication outside the scope of
this Letter.

Any measurement in quantum mechanics can be modeled by a positive
operator-valued measure (POVM) $E(y)$ \cite{holevo,helstrom},
which determines the probability of the observation $y$:
\begin{align}
P(y|g) &= \trace \Bk{E(y)\rho}.
\end{align}
For example, in the direct detection scheme (Fig.~\ref{telescope}(a)),
the two optical modes are brought to interfere at a 50-50 beam
splitter and the photons at the two output ports are counted. It can
be shown by standard quantum optics calculations \cite{sup} that the
POVM $E(n,m)$ for photon counts $n$ and $m$ are
\begin{align}
E(0,0) &= \ket{0,0}\bra{0,0},
\\
E(1,0) &= \frac{1}{2}\bk{\ket{1,0}+e^{-i\delta}\ket{0,1}}
\bk{\bra{1,0}+e^{i\delta}\bra{0,1}},
\\
E(0,1) &= \frac{1}{2}\bk{\ket{1,0}-e^{-i\delta}\ket{0,1}}
\bk{\bra{1,0}-e^{i\delta}\bra{0,1}},
\end{align}
where $\delta$ is an adjustable phase shift on the $b$ mode. The
observation probabilities become
\begin{align}
P(0,0|g) &= 1-\epsilon,
\\
P(1,0|g) &= \frac{\epsilon}{2}\Bk{1+\real(g e^{-i\delta})},
\\
P(0,1|g) &= \frac{\epsilon}{2}\Bk{1-\real (g e^{-i\delta})}.
\end{align}

To evaluate the parameter-estimation capability of a measurement
scheme, consider the Fisher information matrix, defined as
\cite{vantrees}
\begin{align}
F &\equiv 
\sum_y \frac{1}{P(y|g)}D(y|g),
\label{F}
\\
D(y|g) &\equiv \bk{\begin{array}{cc}\Bk{\parti{P(y|g)}{g_1}}^2&
\parti{P(y|g)}{g_1}\parti{P(y|g)}{g_2}\\
\parti{P(y|g)}{g_2}\parti{P(y|g)}{g_1} &
\Bk{\parti{P(y|g)}{g_2}}^2 \end{array}}.
\label{D}
\end{align}
The inverse of the Fisher information matrix provides a lower
Cram\'er-Rao bound to the mean-square estimation error covariance
matrix $\Sigma$ for any unbiased estimate in the form of $\Sigma \ge
F^{-1}$. The eigenvalues of $F$, which must be nonnegative as $F\ge
0$, hence quantify the amounts of independent information obtainable
from the measurement.  In a total observation time interval $T$ over
which the model parameters can be approximated as time-constant,
$M\sim T/\Delta t \sim T\Delta\nu$ measurements can be performed, and
the total Fisher information is $F^{(M)} = MF\sim T\Delta\nu F$. In
the limit of large $M$, the Cram\'er-Rao bound is asymptotically
achievable by maximum-likelihood estimation. This makes the Fisher
information a rigorous metric for comparing the inherent capabilities
of different measurement schemes for parameter estimation.

The Fisher information for direct detection is
\begin{align}
F &= \frac{\epsilon}{1-\real(g e^{-i\delta})^2}
\bk{\begin{array}{cc}\cos^2\delta & \sin\delta\cos\delta\\
\sin\delta\cos\delta & \sin^2\delta\end{array}},
\end{align}
and the eigenvalues of $F$ are
\begin{align}
\lambda_1 &= 0,
&
\lambda_2 &= \frac{\epsilon}{1-\real(g e^{-i\delta})^2}.
\end{align}
The zero eigenvalue corresponds to the absence of information about
the unobservable quadrature $\imag(g e^{-i\delta})$. In practice,
$\delta$ is varied over measurements to retrieve information about
both quadratures of $g$.  The important point to note here is that
$||F|| = \lambda_1+\lambda_2 \ge \epsilon$ if we take the trace
norm. The norm of the total Fisher information for $M$ measurements
becomes $||F^{(M)}|| = M||F|| \ge M\epsilon$, which scales linearly
with the average photon number $M\epsilon$, thereby achieving the
optimal ``shot-noise'' scaling for parameter estimation using
classical states \cite{glm}. Similarly, it is shown in \cite{sup} that
the Fisher information for the shared-entanglement scheme proposed by
Gottesman, Jennewein, and Croke \cite{gottesman} has in theory the
same expression but reduced by a factor of 2.

Both of the aforementioned schemes can be considered as nonlocal
quantum measurements, which require bringing the two modes together
physically or sharing entanglement between the two sites. The physical
nonlocality makes such schemes increasingly challenging to implement
technically as the distance between the two modes increases, primarily
due to accumulating decoherence in the form of optical loss along the
paths \cite{monnier}.  Local measurement schemes, on the other hand,
measure the two modes separately before combining the results via
classical communication (Fig.~\ref{telescope}(b)), and can therefore
be implemented over a much greater distance in principle. To
investigate the general performance of any local measurement, let us
write the observation probability distribution for POVM $E(y)$
explicitly as
\begin{align}
P(y|g) &= (1-\epsilon) E_{00,00}(y) + \frac{\epsilon}{2}
\big[E_{01,01}(y)
\nonumber\\&\quad
+E_{10,10}(y) + 
 2|E_{10,01}(y)|\real(ge^{-i\delta})\big],
\end{align}
where
\begin{align}
E_{n m,n'm'}(y) &\equiv \bra{n,m}E(y)\ket{n',m'}
\end{align}
and $\delta$ is the phase of $E_{10,01}$. To put a bound on the Fisher
information given by Eq.~(\ref{F}), note that
\begin{align}
P(y|g) &\ge (1-\epsilon) E_{00,00}(y),
\label{bound_P}
\end{align}
and the positive-semidefinite matrix 
\begin{align}
D &= \epsilon^2 |E_{10,01}(y)|^2 
\bk{\begin{array}{cc}
\cos^2\delta & \sin\delta\cos\delta\\
\sin\delta\cos\delta & \sin^2\delta\end{array}}
\end{align}
defined in Eq.~(\ref{D}) has a trace norm given by
$\epsilon^2|E_{10,01}(y)|^2$.  Applying the subadditivity property of
matrix norms to Eq.~(\ref{F}) results in an upper bound on $||F||$:
\begin{align}
||F|| &\le \frac{\epsilon^2}{1-\epsilon}
\sum_y \frac{ |E_{10,01}(y)|^2}{E_{00,00}(y)}.
\label{bound}
\end{align}
For generality, I define local measurements as the ones performed
using local operations with classical communication (LOCC), which
permits the measurement at one site to be conditioned upon the
observation at the other site. A necessary condition for a
spatial-LOCC POVM is the positive-partial-transpose condition
$E^{\mathcal T_a}(y) \ge 0$ \cite{terhal}.  By the Cauchy-Schwarz
inequality, $|\bra{1,0}E\ket{0,1}|^2 = |\bra{0,0}E^{\mathcal
  T_a}\ket{1,1}|^2 = |\bra{0,0}\sqrt{E^{\mathcal
    T_a}}\sqrt{E^{\mathcal T_a}}\ket{1,1}|^2 \le \bra{0,0}E^{\mathcal
  T_a}\ket{0,0}\bra{1,1}E^{\mathcal T_a}\ket{1,1}=
\bra{0,0}E\ket{0,0}\bra{1,1}E\ket{1,1}$, or
\begin{align}
|E_{10,01}(y)|^2 &\le E_{00,00}(y) E_{11,11}(y).
\label{ppt}
\end{align}
Combining Eqs.~(\ref{bound}) and (\ref{ppt}), I obtain an
$O(\epsilon^2)$ upper bound on $||F||$:
\begin{align}
||F|| &\le \frac{\epsilon^2}{1-\epsilon}
\sum_y E_{11,11}(y) = \frac{\epsilon^2}{1-\epsilon},
\label{bound2}
\end{align}
where $\sum_y E_{11,11}(y) = 1$ comes from the completeness property
of a POVM. The neglected $O(\epsilon^2)$ term in the density operator
in Eq.~(\ref{rho_number}) contributes an additional $O(\epsilon^2)$
term to $P$ and an $O(\epsilon^3)$ term to $D$, so the Fisher
information would be modified by an $O(\epsilon^3)$ term and the upper
bound in Eq.~(\ref{bound2}) should be rewritten as
\begin{align}
||F|| &\le \epsilon^2 + O(\epsilon^3).
\label{locc_bound}
\end{align}
For $M$ measurements, the bound can be generalized to allow for
adaptive measurements conditioned upon previous observations,
as shown in Ref.~\cite{sup}:
\begin{align}
||F^{(M)}|| &\le M\Bk{\epsilon^2+O(\epsilon^3)}.
\label{locc_boundM}
\end{align}
This upper bound shows that the best Fisher information any
spatiotemporal-LOCC measurement can achieve is still substantially
worse than that of the spatially nonlocal methods ($||F^{(M)}||\sim
M\epsilon$) when $\epsilon \ll 1$. In other words, spatially local
measurements are fundamentally much less efficient than nonlocal
methods in extracting coherence information from weak-thermal-light
interferometry. This general proof is supported by explicit
Fisher-information calculations for heterodyne and homodyne detection
\cite{sup}, signal-to-noise-ratio calculations for direct and
heterodyne detection of the full thermal state given by
Eq.~(\ref{thermal}) \cite{sup}, and the known fact in astronomy that
direct detection performs better than heterodyne detection for high
optical $\nu_0$ \cite{thompson,townes}.  Ref.~\cite{sup} also includes
a discussion of the quantum origin of the nonlocality in terms of the
semiclassical photodetection picture.

Note that the advantage of nonlocal measurements is lost for coherent
states, strong thermal light with $\epsilon \gg 1$ \cite{sup}, or even
the nonclassical single-photon state studied in Ref.~\cite{gottesman}.
For coherent states, $|g| = 1$ and the unknown parameters are the
phases of the two optical modes in a product of coherent states, in
which case it can easily be shown that nonlocal measurements are not
necessary, analogous to the case of single-parameter phase estimation
with a product state \cite{glm_prl}. For strong thermal light with
$\epsilon \gg 1$, calculations in Ref.~\cite{sup} show that the
performances of direct detection and heterodyne detection converge and
suggest that the noise in this regime is dominated by the thermal
statistics of the source rather than the detection statistics. The
single-photon state studied in Ref.~\cite{gottesman} can also be
analyzed using the formalism here by omitting $O(\epsilon^2)$ terms
and then putting $\epsilon = 1$, resulting in comparable performances
for local and nonlocal measurements.

The peculiar existence of quantum nonlocality for weak thermal light,
as a property of bipartite measurements applied to certain separable
states, can be regarded as a dual of Einstein-Podolsky-Rosen
entanglement \cite{holevo,epr}, a property of bipartite states that
can produce higher correlations in certain separable measurements. In
the context of quantum communication theory, it is well known that
nonlocal measurements can extract more information from states with no
entanglement \cite{holevo,peres,lloyd}; the result here provides a
striking example in which the same type of quantum nonlocality readily
exists for observers extracting information from nature.

For practical applications, the result here demonstrates the
fundamental advantage of nonlocal quantum measurements for
weak-thermal-light interferometry and may have further implications
for optical imaging systems, such as compound-eye imaging and
fluorescence microscopy \cite{brady}. The shared-entanglement proposal
in Ref.~\cite{gottesman} requires a path-entangled single-photon
source and quantum repeaters, both of which are unlikely to become
feasible in the near future, but standard linear optics can also
perform nonlocal measurements by coherently processing multiple
optical modes before detection, provided that optical loss can be
minimized. In the short term, the result here thus motivates the
development of low-loss coherent optical information devices, such as
photonic crystal fibers and integrated photonics, for thermal-light
interferometry and imaging \cite{monnier,brady,psaltis}.

Accurate coherence information can be obtained only in the limit of
many collected photons. This corresponds to measurements of many
copies of the quantum state. A more general quantum measurement
strategy than the ones considered here involves joint quantum
operations on the multiple copies before measurements. This kind of
temporal nonlocality is not needed for parameter estimation when
spatially nonlocal measurements can be performed \cite{glm}.  It
remains an interesting open question whether coherent temporally
nonlocal strategies can offer any significant advantage when one is
restricted to spatially local measurements.  Other potential
generalizations include time-varying parameters and the estimation of
temporal coherence for spectroscopy in addition to spatial
coherence. One must then take into account the dynamics of the source
and colored noise, which can be analyzed using the quantum waveform
estimation framework developed in Refs.~\cite{tsang}.

This material is based on work supported in part by the Singapore
National Research Foundation under NRF Award No.~NRF-NRFF2011-07, NSF
Grants No.~PHY-0903953, No.~PHY-1005540, and ONR Grant
No.~N00014-11-1-0082.  Discussions with Daniel Gottesman, Carlton
Caves, Howard Wiseman, Alexander Lvovsky, Christoph Simon, Mohan
Sarovar, Alexander Tacla, Jiang Zhang, Matthias Lang, Shashank Pandey,
and Stefano Pirandola are gratefully acknowledged.

\appendix

\section*{Supplementary Material}
This Supplementary Material contains supportive calculations and
discussions that complement the main text. Section~\ref{dd} derives
the positive operator-valued measure (POVM) for direct detection,
Sec.~\ref{gjc} derives the POVM for shared-entanglement interferometry
and the resulting Fisher information, Sec.~\ref{adaptive} calculates a
bound on the total Fisher information for multiple adaptive
measurements, Secs.~\ref{hetero} and \ref{homo} calculate the Fisher
information for heterodyne and homodyne detection, Sec.~\ref{strong}
investigates the performances of direct and heterodyne detection for
strong thermal light, and Sec.~\ref{semi} discusses the origin of the
quantum nonlocality in terms of the semiclassical photodetection
picture. The reference list at the end is identical to the one in the
main text for easier cross-referencing.

\section{\label{dd}POVM for direct detection}
In direct detection, the optical modes $a$ and $b$ are combined by a
beam splitter and photon-counting is performed at the two output
ports. Let $U$ be the unitary operator that corresponds to the
operation of the beam splitter on the bipartite quantum state $\rho$.
The observation probability distribution is
\begin{align}
P(n,m|g) &= \bra{n,m}U\rho U^\dagger\ket{n,m},
\end{align}
where $\ket{n,m}$ is a Fock state. We can then write the POVM as 
\begin{align}
E(n,m) &= U^\dagger \ket{n,m}\bra{n,m} U,
\end{align}
which propagates the Fock-state projection back to the time when the
state of light is $\rho$.

With at most one photon in the quantum state, we are interested
in $(n,m) = (0,0), (1,0), (0,1)$ only. Applying the unitary 
to the Fock states,
\begin{align}
U^\dagger\ket{0,0} &= \ket{0,0},
\\
U^\dagger \ket{1,0} &= U^\dagger a^\dagger U U^\dagger\ket{0,0}
= U^\dagger a^\dagger U\ket{0,0}
\\
&= \frac{1}{\sqrt{2}}\bk{a^\dagger + e^{-i\delta} b^\dagger}
\ket{0,0}
\\
&= \frac{1}{\sqrt{2}}\bk{\ket{1,0}+e^{-i\delta} \ket{0,1}},
\\
U^\dagger\ket{0,1} &= 
\frac{1}{\sqrt{2}}\bk{\ket{1,0}-e^{-i\delta} \ket{0,1}}.
\end{align}
The POVM is hence
\begin{align}
E(0,0) &= \ket{0,0}\bra{0,0},
\\
E(1,0) &= \frac{1}{2}\bk{\ket{1,0}+e^{-i\delta} \ket{0,1}}
\bk{\bra{1,0}+e^{i\delta} \bra{0,1}},
\\
E(0,1) &= \frac{1}{2}\bk{\ket{1,0}-e^{-i\delta} \ket{0,1}}
\bk{\bra{1,0}-e^{i\delta} \bra{0,1}}.
\end{align}

\section{\label{gjc}Shared-entanglement interferometry}
Assuming an entangled ancilla in two modes $c$ and $d$ given by
\begin{align}
\ket{\delta} &\equiv \frac{1}{\sqrt{2}}
\bk{\ket{0,1}_{c,d} + e^{i\delta}\ket{1,0}_{c,d}},
\end{align}
with each mode sent to the sites of $a$ and $b$ modes for separate
interference measurements \cite{gottesman}, the POVM for photon counts
$(n,m,n',m')$ is
\begin{align}
E(n,m,n',m') &= \bra{\delta}U_{ac}^\dagger\otimes U_{bd}^\dagger\ket{n,m,n',m'}
\bra{n,m,n',m'} U_{ac}\otimes U_{bd}\ket{\delta},
\end{align}
where $U_{ac}$ denotes the beam-splitting unitary on modes $a$ and $c$
and $U_{bd}$ denotes the same unitary on modes $b$ and $d$.  The
calculation is more involved but similar to the one for direct
detection. The final result is
\begin{align}
E(y_0) &= \ket{0,0}\bra{0,0},
\\
E(y_1) &= \frac{1}{2}\ket{0,1}\bra{0,1},
\\
E(y_2) &= \frac{1}{2}\ket{1,0}\bra{1,0},
\\
E(y_3) &= \frac{1}{4}\bk{\ket{1,0}+e^{-i\delta}\ket{0,1}}
\bk{\bra{1,0}+e^{i\delta} \bra{0,1}},
\\
E(y_4) &= \frac{1}{4}\bk{\ket{1,0}-e^{-i\delta} \ket{0,1}}
\bk{\bra{1,0}-e^{i\delta} \bra{0,1}},
\end{align}
where each $y_j$ corresponds to a set of $(n,m,n',m')$ that produce
the same POVM. When applied to the quantum state $\rho$, only
observations $y_3$ and $y_4$ contribute to the Fisher information
about $g$.  Since $E(y_3) = E(1,0)/2$ and $E(y_4) = E(0,1)/2$,
the Fisher information for shared-entanglement interferometry is
simply that for direct detection reduced by a factor of 2:
\begin{align}
F &= \frac{\epsilon}{2[1-\real(g e^{-i\delta})^2]}
\bk{\begin{array}{cc}\cos^2\delta & \sin\delta\cos\delta\\
\sin\delta\cos\delta & \sin^2\delta\end{array}}.
\end{align}

\section{\label{adaptive}Adaptive measurements}
For $M$ measurements, the joint observation probability distribution
can be written as
\begin{align}
P(y_M,\dots,y_1|g) &= P(y_M|g,y_{M-1},\dots,y_1)
P(y_{M-1},\dots,y_1|g).
\end{align}
Each element of the total Fisher information matrix for $M$
measurements, using an alternate form of the Fisher matrix \cite{vantrees},
becomes
\begin{align}
F_{jk}^{(M)} &\equiv -\sum_{y_1,\dots,y_M} P(y_M,\dots,y_1|g)
\parti{^2}{g_j\partial g_k}
\ln P(y_M,\dots,y_1|g)
\\
&= -\sum_{y_1,\dots,y_M} P(y_M|g,y_{M-1},\dots,y_1)P(y_{M-1},\dots,y_1|g)
\nonumber\\&\quad\times
\Bk{\parti{^2}{g_j\partial g_k}\ln P(y_{M}|g,y_{M-1},\dots,y_1)
+\parti{^2}{g_j\partial g_k}\ln P(y_{M-1},\dots,y_1|g)}
\\
&= \sum_{y_1,\dots,y_{M-1}} P(y_{M-1},\dots,y_1|g)F_{M jk}(y_{M-1},\dots,y_1)
+F_{jk}^{(M-1)},
\end{align}
where $F_M$ denotes the conditional Fisher information of the $M$th
measurement:
\begin{align}
F_{M jk}(y_{M-1},\dots,y_1) &\equiv 
-\sum_{y_M}P(y_M|g,y_{M-1},\dots,y_1)
\parti{^2}{g_j\partial g_k}\ln P(y_{M}|g,y_{M-1},\dots,y_1).
\end{align}
Applying the subadditivity property of matrix norms,
\begin{align}
||F^{(M)}|| &\le \sum_{y_1,\dots,y_{M-1}} P(y_{M-1},\dots,y_1|g)||F_M||
+||F^{(M-1)}||
\\
&\le \max_{y_1,\dots,y_{M-1}} ||F_M||+||F^{(M-1)}||,
\end{align}
and by induction, 
\begin{align}
||F^{(M)}|| &\le \sum_{m=1}^M \max_{y_1,\dots,y_{m-1}} ||F_m||.
\end{align}
This proves that the norm of the total Fisher information cannot
exceed the sum of the maximized single-measurement values.

For the $m$th quantum measurement with outcome $y_m$ conditioned upon
previous observations, we can write
\begin{align}
P(y_m|g,y_{m-1},\dots,y_1)
&= \trace\Bk{E(y_m|y_{m-1},\dots,y_1)\rho}.
\end{align}
This means that the bound given by Eq.~(23) in the main text for
spatial-LOCC measurements in the case of $\epsilon \ll 1$ is also
applicable to $||F_m||$:
\begin{align}
||F_{m}|| &\le \epsilon^2 + O(\epsilon^3).
\end{align}
The total Fisher information is hence bounded by
\begin{align}
||F^{(M)}|| &\le M\Bk{\epsilon^2+O(\epsilon^3)},
\end{align}
which generalizes the bound to the case of multiple
spatiotemporal-LOCC measurements and proves that no adaptive strategy
can improve the scaling $||F^{(M)}|| \sim M\epsilon^2$.

\section{\label{hetero}Heterodyne detection}

The POVM for heterodyne detection is \cite{helstrom}
\begin{align}
E(\mu,\nu) &= \frac{1}{\pi^2}\ket{\mu,\nu}\bra{\mu,\nu},
\end{align}
where $\ket{\mu,\nu}$ is a coherent state and the normalization is
$\int d^2\mu d^2\nu E(\mu,\nu) = I$, the identity operator. The
relevant POVM matrix elements are
\begin{align}
E_{00,00}(\mu,\nu) &\equiv \frac{1}{\pi^2}\abs{\avg{0,0|\mu,\nu}}^2
\\
&= \frac{1}{\pi^2}\exp\bk{-|\mu|^2-|\nu|^2},
\\
E_{01,01}(\mu,\nu) &\equiv \frac{1}{\pi^2}\abs{\avg{0,1|\mu,\nu}}^2
\\
&= \frac{1}{\pi^2}\exp\bk{-|\mu|^2-|\nu|^2}|\nu|^2,
\\
E_{10,10}(\mu,\nu) &\equiv \frac{1}{\pi^2}\abs{\avg{1,0|\mu,\nu}}^2
\\
&=\frac{1}{\pi^2}\exp\bk{-|\mu|^2-|\nu|^2}|\mu|^2,
\\
E_{01,10}(\mu,\nu) &\equiv \frac{1}{\pi^2}\avg{0,1|\mu,\nu}
\avg{\mu,\nu|1,0}
\\
&= \frac{1}{\pi^2}\exp\bk{-|\mu|^2-|\nu|^2}\mu^* \nu.
\end{align}
The Fisher information is hence
\begin{align}
F &= \frac{\epsilon^2}{2}\bk{\begin{array}{cc}1&0\\0&1\end{array}}
+O(\epsilon^3),
\\
||F|| &= \epsilon^2 + O(\epsilon^3).
\end{align}
This shows that the performance of heterodyne detection is already the
optimum allowed by quantum mechanics for any local measurement
according to the bound given by Eq.~(23) in the main text.

\section{\label{homo}Homodyne detection}
For homodyne detection,
\begin{align}
E(x,y) &= \ket{x,y}\bra{x,y},
\end{align}
where $\ket{x,y}$ is a quadrature eigenstate:
\begin{align}
\frac{1}{\sqrt{2}} \bk{a e^{-i\delta_a} + a^\dagger e^{i\delta_a}}
\ket{x,y} &= x\ket{x,y},
\\
\frac{1}{\sqrt{2}} \bk{b e^{-i\delta_b} + b^\dagger e^{i\delta_b}}
\ket{x,y} &= y\ket{x,y},
\end{align}
$\delta_a$ and $\delta_b$ are local-oscillator phases, and the
normalization is $\int dx dy E(x,y) = I$.  The relevant POVM elements
are
\begin{align}
E_{00,00}(x,y) &= \frac{1}{\pi}\exp\bk{-x^2-y^2},
\\
E_{01,01}(x,y) &= \frac{2}{\pi}
\exp\bk{-x^2-y^2}y^2,
\\
E_{10,10}(x,y) &= \frac{2}{\pi}
\exp\bk{-x^2-y^2}x^2,
\\
E_{10,01}(x,y) &= \frac{2}{\pi} e^{i\delta}
\exp\bk{-x^2-y^2}x y,
\end{align}
where $\delta \equiv \delta_a-\delta_b$. The Fisher information
becomes
\begin{align}
F &= \epsilon^2\bk{\begin{array}{cc}
\cos^2\delta & \sin\delta\cos\delta\\
\sin\delta\cos\delta & \sin^2\delta
\end{array}} + O(\epsilon^3),
\\
||F|| &= \epsilon^2+O(\epsilon^3).
\end{align}
Homodyne detection is also able to saturate the bound given by
Eq.~(23) in the main text, but each measurement gives information
about only one quadrature of $g$ and $\delta$ should be varied over
measurements to estimate both quadratures.

Qualitatively, the inferior Fisher information for heterodyne and
homodyne detection can be attributed to the non-zero vacuum
fluctuations even when no photon is coming in to provide information
about the unknown parameters. Nonlocal measurements are able to
perfectly discriminate against this case and discard the useless
observations, but heterodyne or homodyne detection is unable to do so
and forced to include vacuum fluctuations as potentially useful
observations, resulting in a substantially worse estimation accuracy
in the long run.

\section{\label{strong}Thermal light with arbitrary $\epsilon$}
For $\epsilon \gtrsim 1$, it is necessary to use the full thermal
state given by Eq.~(1) in the main text. First consider the
observation probability density of heterodyne detection:
\begin{align}
P(\mu,\nu|g) &= \int d^2\alpha d^2\beta
\Pi(\mu,\nu|\alpha,\beta)\Phi(\alpha,\beta|g),
\label{thermal_hetero}
\\
\Pi(\mu,\nu|\alpha,\beta) &\equiv 
\bra{\alpha,\beta}E(\mu,\nu)\ket{\alpha,\beta}
\\
&= \frac{1}{\pi^2}
\exp\bk{-|\mu-\alpha|^2-|\nu-\beta|^2}.
\end{align}
We can interpret these expressions using a semiclassical
photodetection picture \cite{mandel}: The heterodyne detection
statistics obey $\Pi(\mu,\nu|\alpha,\beta)$ for given classical fields
$(\alpha,\beta)$, but the fields from the source also have a
statistical distribution given by $\Phi(\alpha,\beta)$, so the
marginal observation density is taken to be $\Pi$ averaged over
$\Phi$.  The resulting convolution of the two Gaussians can be
calculated analytically and given by
\begin{align}
P(\mu,\nu|g) &= 
\frac{1}{\pi^2\det \Gamma'}
\exp\Bk{-\bk{\begin{array}{cc}\mu^* &\nu^*\end{array}}
\Gamma'^{-1}\bk{\begin{array}{c}\mu\\\nu\end{array}}},
\end{align}
where the new covariance matrix $\Gamma'$ is
\begin{align}
\Gamma' &= \Gamma + \bk{\begin{array}{cc}1&0\\0&1\end{array}}
=\bk{\begin{array}{cc}\epsilon/2+1 & \epsilon g/2\\
\epsilon g^*/2 & \epsilon/2 + 1\end{array}}.
\end{align}
The factors of $1$ come from $\Pi$ and represent detection noise.
When $\epsilon \gg 1$, $\Pi$ is much sharper than $\Phi$, so the
convolution essentially reproduces $\Phi$ as the marginal observation
density and $P(\mu,\nu|g)\approx \Phi(\mu,\nu|g)$. In other words, the
inherent thermal noise from the source overwhelms the heterodyne
detection noise in the $\epsilon \gg 1$ regime.

To estimate the performance of heterodyne detection, we can consider
the signal-to-noise ratio (SNR) \cite{thompson,townes}.  If we take $\mu\nu^*$ as the
output signal, $\avg{\mu\nu^*} = \epsilon g/2$ directly gives $g$ on
average, and the signal energy is
\begin{align}
S &\equiv \abs{\Avg{\mu\nu^*}}^2 = \frac{\epsilon^2|g|^2}{4}.
\end{align}
The noise energy is
\begin{align}
N  &\equiv \Avg{|\mu\nu^*|^2}-S.
\end{align}
The fourth-order field statistics can be computed with
the help of the matrix $G \equiv \Gamma'^{-1}$:
\begin{align}
\det G &= G_{aa}G_{bb}-G_{ab}G_{ba},
\\
\Avg{|\mu\nu^*|^2}
&= \det G\parti{^2}{G_{aa}\partial G_{bb}}\frac{1}{\det G}
\\
&= \frac{2G_{aa}G_{bb}}{\det G^2}-\frac{1}{\det G}
\\
&= \bk{1+\frac{\epsilon}{2}}^2+\frac{\epsilon^2|g|^2}{4},
\\
N &= \bk{1+\frac{\epsilon}{2}}^2.
\end{align}
The SNR is hence
\begin{align}
\frac{S}{N} &= \frac{\epsilon^2|g|^2}{(2+\epsilon)^2}.
\end{align}
For $\epsilon \ll 1$, $S/N\approx \epsilon^2|g|^2/4$, but for
$\epsilon \gg 1$, the SNR saturates to $S/N\to |g|^2$ and becomes
independent of $\epsilon$.

For direct detection,
\begin{align}
P(n,m|g) &= \int d^2\alpha d^2\beta
\Pi(n,m|\alpha,\beta)\Phi(\alpha,\beta|g),
\\
\Pi(n,m|\alpha,\beta) &\equiv 
\bra{\alpha,\beta}E(n,m)\ket{\alpha,\beta}
=\abs{\avg{n,m|u,v}}^2,
\\
\bk{\begin{array}{c}u\\v\end{array}} &\equiv 
V\bk{\begin{array}{c}\alpha\\\beta\end{array}},
\quad
V \equiv \frac{1}{\sqrt{2}}\bk{\begin{array}{cc}1&e^{i\delta}\\
1 & -e^{i\delta}\end{array}}.
\end{align}
Changing the integration variables to $(u,v)$,
\begin{align}
P(n,m|g) &=  
\int d^2u d^2v
\Pi'(n,m|u,v)\Phi'(u,v|g),
\\
\Pi'(n,m|u,v)&\equiv \exp(-|u|^2)\frac{|u|^{2n}}{n!}
\exp(-|v|^2)\frac{|v|^{2m}}{m!},
\\
\Phi'(u,v|g) &\equiv 
\frac{\det K}{\pi^2}
\exp\Bk{-\bk{\begin{array}{cc}u^* &v^*\end{array}}
K\bk{\begin{array}{c}u\\ v\end{array}}},
\\
K &\equiv V\Gamma^{-1}V^\dagger
= \frac{2}{\epsilon(1-|g|^2)}
\bk{\begin{array}{cc}1-\real(ge^{-i\delta}) & i\imag(ge^{-i\delta})\\
-i\imag(ge^{-i\delta}) & 1+\real(ge^{-i\delta}) \end{array}},
\\
\det K &= K_{aa}K_{bb}-K_{ab}K_{ba} = \frac{4}{\epsilon^2(1-|g|^2)}.
\end{align}
The averaging of a Poissonian with a Gaussian is difficult to
calculate exactly. For $\epsilon \gg 1$, however, the photon counts
$(n,m)$ should be large most of the time, and $P(n,m|g)$ may be
approximated by a Gaussian.  Let us therefore focus on the first and
second moments of $(n,m)$ for $P(n,m|g)$. The first moment of $n$ is
\begin{align}
\Avg{n} &= \sum_{n,m} n P(n,m|g)
\\
&= \int d^2u d^2v |u|^2\Phi'(u,v|g)  \\
&= -\det K\parti{}{K_{aa}}\frac{1}{\det K}.
\end{align}
Similarly,
\begin{align}
\Avg{m} &= -\det K\parti{}{K_{bb}}\frac{1}{\det K},
\\
\Avg{n^2} &=  \Avg{n} + \det K\partit{}{K_{aa}} \frac{1}{\det K},
\\
\Avg{m^2} &=\Avg{m} + \det K\partit{}{K_{bb}} \frac{1}{\det K},
\\
\Avg{nm} &= \det K\parti{^2}{K_{aa}\partial K_{bb}} \frac{1}{\det K}.
\end{align}
This gives
\begin{align}
\Avg{n} &= \frac{\epsilon}{2}\Bk{1+\real(ge^{-i\delta})},
\\
\Avg{m} &= \frac{\epsilon}{2}\Bk{1-\real(ge^{-i\delta})},
\\
\Avg{\Delta n^2} &= \Avg{n} + \Avg{n}^2,
\\
\Avg{\Delta m^2} &= \Avg{m} + \Avg{m}^2,
\\
\Avg{\Delta n\Delta m} 
&= \frac{\epsilon^2}{4}\imag(ge^{-i\delta})^2.
\end{align}
A behavior similar to the case of heterodyne detection can be seen
here. For $\avg{n}, \avg{m} \sim \epsilon \gg 1$, the noise
covariances scale as $\epsilon^2$ rather than $\epsilon$, indicating
that the source thermal noise also overwhelms the Poissonian detection
noise.

Since the observation statistics are expected to be approximately
Gaussian for $\epsilon \gg 1$, we can similarly consider the SNR as a
performance metric. Taking the output signal as $n-m$, the average of
which gives $\avg{n-m} = \epsilon \real(ge^{-i\delta})$, a quadrature
of $g$, the signal energy is
\begin{align}
S &= \Avg{n-m}^2 = \epsilon^2 \real(ge^{-i\delta})^2,
\end{align}
and the noise energy is
\begin{align}
N &= \Avg{(n-m)^2}-S
\\
&= \Avg{\Delta n^2} + \Avg{\Delta m^2} - 2\Avg{\Delta n\Delta m}
\\
&= \epsilon+\frac{\epsilon^2}{2}\Bk{1+ \real(ge^{-i\delta})^2
-\imag(ge^{-i\delta})^2}.
\end{align}
If we perform two measurements, one with $\delta = \delta_1$ and one
with $\delta = \delta_1+\pi/2$ to measure the other quadrature of $g$,
the average signal and noise energies per measurement becomes
\begin{align}
\bar S  &= \frac{\epsilon^2|g|^2}{2},
\\
\bar N &= \epsilon + \frac{\epsilon^2}{2},
\end{align}
and the average SNR is
\begin{align}
\frac{\bar S}{\bar N} &= \frac{\epsilon|g|^2}{2+\epsilon}.
\end{align}
For $\epsilon \gg 1$, the SNR saturates to $|g|^2$, just like the SNR
of heterodyne detection, suggesting that the SNR is dominated by
source thermal noise regardless of the detection method and nonlocal
measurements do not have an advantage when $\epsilon \gg 1$.

For $\epsilon \ll 1$, the SNR is still a valid performance metric for
a large number of measurements, in which case the statistics become
approximately Gaussian by the central limit theorem and averaging $M$
observations improves the final SNR by a factor of $M$. The
direct-detection SNR is $\approx M\epsilon|g|^2/2$ and significantly
better than the heterodyne SNR $\approx M\epsilon^2|g|^2/4$, a fact
well known in astronomy \cite{thompson,townes} and rigorously
generalized in this paper.

\section{\label{semi}Quantum origin of measurement nonlocality in the
  semiclassical photodetection picture}
One may well wonder where quantum mechanics comes in, if both $\Phi$
and $\Pi$ are nonnegative and the whole problem obeys classical
statistics in the semiclassical photodetection picture. The answer
lies in the fact that the likelihood function
\begin{align}
\Pi(y|\alpha,\beta) &\equiv \bra{\alpha,\beta}E(y)\ket{\alpha,\beta}
\end{align}
cannot be an arbitrarily sharp probability distribution in quantum
mechanics.  It is the Husimi representation \cite{mandel}, more commonly applied
to a quantum state but here to a POVM.

If we regard $\Pi(y|\alpha,\beta)$ as a likelihood function of
$(\alpha,\beta)$ for a given observation $y$, the sharpness of
$\Pi(y|\alpha,\beta)$ with respect to $(\alpha,\beta)$ in phase space
characterizes the amount of information about $(\alpha,\beta)$
contained in the observation $y$. The Husimi representation
has a maximum magnitude and a finite variance for each quadrature,
which means that there is a limited amount of information about the
fields that an observation can provide.

The information of mutual coherence lies only in the nonlocal
second-order field correlation $\alpha\beta^*$ for thermal light, the
first-order mean fields of which are zero. If $E(y)$ corresponds to a
local measurement and is separable into $E_a(y)\otimes E_b(y)$, the
sharpness of $\Pi(y|\alpha,\beta) = \Pi_a(y|\alpha) \Pi_b(y|\beta)$
with respect to $\alpha\beta^*$ would be more limited than that
allowed by nonlocal measurements, meaning that local measurements
extract less information about the coherence than nonlocal
measurements. In this sense, the measurement nonlocality can be
regarded as a dual of Einstein-Podolsky-Rosen entanglement
\cite{holevo,epr}; the former is a property of bipartite measurements
that can extract more information from certain separable states and
the latter a property of bipartite states that can produce higher
correlations in certain separable measurements.

\end{document}